\documentclass[aps,prd,twocolumn,superscriptaddress,showpacs,showkeys]{revtex4}

\usepackage{epsfig,epsf}
\usepackage{amsmath}
\usepackage{amsthm}
\usepackage{amsfonts}
\usepackage{amssymb}
\usepackage{dsfont}

\usepackage{multirow}

\usepackage{slashed}

\usepackage[active]{srcltx}
\usepackage{psfrag}

\newcommand{\be}{\begin{equation}}
\newcommand{\ee}{\end{equation}}
\newcommand{\ba}{\begin{eqnarray}}
\newcommand{\ea}{\end{eqnarray}}
\newcommand{\baa}{\begin{eqnarray*}}
\newcommand{\btab}{\begin{tabular}}
\newcommand{\etab}{\end{tabular}}
\newcommand{\eaa}{\end{eqnarray*}}

\def\inbar{\,\vrule height1.5ex width.4pt depth0pt}
\def\IC{\relax\hbox{$\inbar\kern-.3em{\rm C}$}}
\def\IZ{\relax{\hbox{\cmss Z\kern-.4em Z}}}
\def\IR{{\hbox{{\rm I}\kern-.2em\hbox{\rm R}}}}

\def\IP{{\hbox{{\rm I}\kern-.2em\hbox{\rm P}}}}
\def\II{\hbox{{1}\kern-.25em\hbox{l}}}

\begin{document}

\title{
BELLE Data on the $\pi^0\gamma^*\gamma$ Form Factor: A Game Changer?
}


\date{\today}

\author{S.S.~Agaev}
\affiliation{Institut f\"ur Theoretische Physik, Universit\"at
   Regensburg, D-93040 Regensburg, Germany}
\affiliation{Institute for Physical Problems, Baku State University,
 Az--1148 Baku, Azerbaijan}
\author{V.M.~Braun}
\affiliation{Institut f\"ur Theoretische Physik, Universit\"at
   Regensburg, D-93040 Regensburg, Germany}
\author{N.~Offen}
\affiliation{Institut f\"ur Theoretische Physik, Universit\"at
   Regensburg, D-93040 Regensburg, Germany}
\author{F.A.~Porkert}
\affiliation{Institut f\"ur Theoretische Physik, Universit\"at
   Regensburg, D-93040 Regensburg, Germany}

\begin{abstract}
We extend our analysis \cite{I} of the $\pi^0\gamma^*\gamma$ form factor
by including a comparison with the new BELLE data~\cite{Uehara:2012ag}.
The necessity of new precision measurements in a broad interval 
of momentum transfers is emphasized.
\end{abstract}



\maketitle


%
%

In this note we analyze the impact of the new
BELLE data~\cite{Uehara:2012ag} on the theoretical status of the $\pi^0\gamma^*\gamma$ 
form factor in QCD and the constraints on the pion distribution amplitude (DA).
It has been a hot subject over the last several years, with 
a strong scaling violation suggested by the BaBar measurements~\cite{BABAR} 
fuelling a flurry of theoretical activity.
 
This analysis has to be viewed as an addendum to our work Ref.~\cite{I} 
where we have given an updated NLO analysis of the form factor 
in the framework of QCD collinear factorization, taking into account higher-twist
corrections up to twist-six and also soft nonperturbative 
corrections estimated using dispersion relations and duality.
In this note we follow Ref.~\cite{I} closely both for the formalism and notations;  
We also refer to~\cite{I} for the complete list of relevant references.

\begin{figure}[t]
\centerline{
\begin{picture}(210,140)(0,0)
\put(-5,0){\epsfxsize7.8cm\epsffile{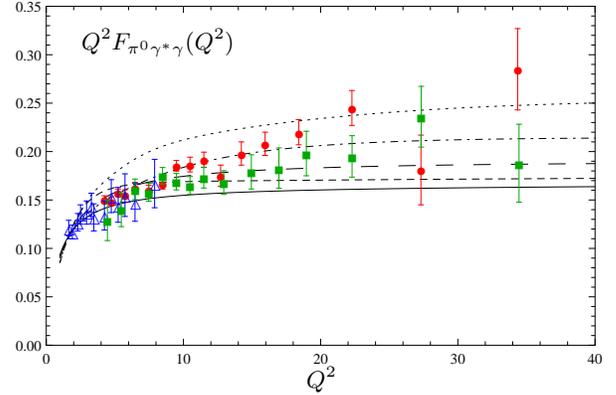}}
\put(105,-8){$Q^2$}
\put(20,120){$Q^2F_{\pi^0\gamma^*\gamma}(Q^2)$}
\end{picture}
}
\caption{
The pion transition form factor for the 
``asymptotic'' (solid line),
``BMS''\cite{BMS} (short dashes), 
``holographic'' (long dashes), 
  ``model~II'' of Ref.~\cite{I} (dash-dotted)
and ``flat'' (dots) pion DA.
The experimental data are from~\cite{Uehara:2012ag} 
(squares), \cite{BABAR} (circles)
and~\cite{Gronberg:1997fj} (open triangles).
}
\label{fig1:SUMMARY}
\end{figure}

A compilation of the experimental data by BELLE, BaBar and CLEO 
collaborations \cite{Uehara:2012ag,BABAR,Gronberg:1997fj} is shown 
in Fig.~\ref{fig1:SUMMARY} together with QCD calculations using several 
models of the pion DA: asymptotic $\phi^{\rm as}_\pi(x) = 6x(1-x)$ (solid line), 
BMS model~\cite{BMS} (short dashes), 
``holographic'' model $\phi^{\rm hol}_\pi(x) = (8/\pi)\sqrt{6x(1-x)}$~\cite{Brodsky:2007hb} 
(long dashes), a model~II of Ref.~\cite{I} (dash-dotted)  
and ``flat'' $\phi^{\rm flat}_\pi(x) = 1$~\cite{Radyushkin:2009zg}
 (dots). 
One sees that the BELLE data are systematically lower than those of BaBar in a broad range of $Q^2$ 
and can be described e.g. by the ``holographic'' model. The quality of the fit is
$\chi^2 = 0.64$ (per data point) adding theoretical uncertainties and experimental errors 
in quadrature. 

\begin{figure}[t]
\centerline{
\begin{picture}(210,140)(0,0)
\put(-5,0){\epsfxsize7.8cm\epsffile{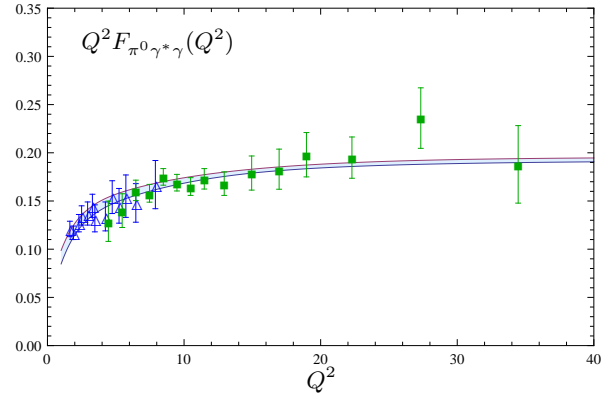}}
\put(105,-8){$Q^2$}
\put(20,120){$Q^2F_{\pi^0\gamma^*\gamma}(Q^2)$}
\end{picture}
}
\caption{
The pion transition form factor for the 
model of the pion DA described in the text.
The estimated theoretical uncertainty is 
shown by the shaded area.  
The experimental data are from~\cite{Uehara:2012ag} 
(squares)
and~\cite{Gronberg:1997fj} (open triangles).
        }
\label{fig2:BELLE}
\end{figure}

\begin{figure}[t]
\begin{center}
\begin{picture}(210,140)(0,0)
\put(0,0){\epsfxsize7.3cm\epsffile{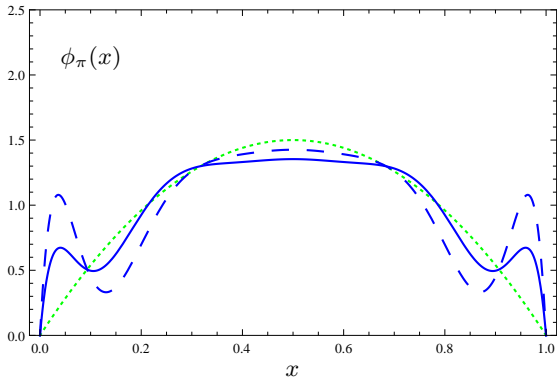}}
\put(105,-8){$x$}
\put(20,110){$\phi_\pi(x)$}
\end{picture}
\end{center}
\caption{The model of the pion DA at the scale 1 GeV
used in Fig.~\ref{fig2:BELLE} (solid line)
compared to the model~II of Ref.~\cite{I} 
shown by the dash-dotted curve in Fig.~\ref{fig1:SUMMARY}
(dashed)
and the asymptotic DA (dotted). 
}
\label{fig3:DAmodels}
\end{figure}

One can try to quantify the difference in the pion DAs which are required to describe 
these two data sets. To this end we take the model~II of Ref.~\cite{I} as an example 
(dash-dotted curve in Fig.~\ref{fig1:SUMMARY}), and modify the parameters 
in order to obtain a good fit for BELLE data instead of BaBar.  
We end up with the following values of the Gegenbauer coefficients at the scale 1 GeV:
\begin{eqnarray}
  a_2 \,=\, 0.10\,(0.14)\,, &\qquad& 
  a_4 \,=\, 0.10\,(0.23)\,,\nonumber\\
  a_6 \,=\, 0.10\,(0.18)\,, &\qquad&
  a_8 \,=\, 0.034\,(0.05)\,,   
\label{model_IV}
\end{eqnarray} 
where the numbers in parenthesis correspond to the model~II of 
Ref.~\cite{I} and, by design, describe the BaBar data~\cite{BABAR}.
The transition form factor calculated using the pion DA defined by Eq.~(\ref{model_IV}) is shown 
in Fig.~\ref{fig2:BELLE} together with the estimated theoretical uncertainty.
It is very similar to the ``holographic'' model, apart from being somewhat lower at small $Q^2$
and provides a better overall fit in this region. 
Adding the theoretical uncertainty and the experimental errors in quadrature, we obtain 
for our new model $\chi^2 = 0.44$ (per BELLE data point). For comparison, we get $\chi^2 = 1.06$ for the 
description of the BaBar data in the model~II of Ref.~\cite{I}.   
These two models for the pion DA, describing alternatively either BELLE or BaBar data, are compared
with each other in Fig.~\ref{fig3:DAmodels}. It is seen that the BaBar measurements imply an 
enhancement of the DA in the end-point region, but apart from that the differences are minor.
This is not surprising since the overall discrepancy between the BELLE and BaBar data sets is within 
1.5-2 standard deviations. 

We conclude that more precise form factor measurements for both small and large momentum transfers 
are needed in order to arrive at a definite conclusion.
Let us  discuss the role of different regions.

\subsubsection*{1.~~moderate $Q^2\sim 2-6$~GeV$^2$}

The QCD predictions in the $Q^2\sim 2-6$~GeV$^2$ rely rather heavily on the 
duality assumption for the treatment of soft contributions and 
also higher-twist corrections appear to be significant. We estimate an irreducible
theoretical uncertainty for this region (for the NLO calculation for a given DA) as $\pm5\%$
but the accuracy of the duality approximation itself is difficult to quantify. 
It has to be tested. A sensitivity to the pion DA in this region is, on the other hand, 
mostly limited to the second-order coefficient $a_2$ in the Gegenbauer expansion 
which can be calculated rather precisely on the lattice.  
With this input, comparing the calculations of the $\pi^0\gamma^*\gamma$ 
transition form factor with the data one will be able to get a much better understanding 
of the theoretical accuracy than presently available. This will be very important in a broader 
context, e.g. for QCD calculations of weak decays of B-mesons which involve similar relatively low scales.
Especially with new data from the BES III collaboration who plan a significant improvement 
over the Cleo, BaBar and Belle results in this region~\cite{Unverzagt:2012zz}
there can be quantitative progress.

\subsubsection*{2.~~intermediate  $Q^2\sim 10$~GeV$^2$}

The existing three data points around $Q^2=10$~GeV$^2$ have small error bars for both BaBar and BELLE,
so that also the difference between the two experiments in this region is the most statistically significant. 
The large values for $Q^2 F_{\pi^0\gamma^*\gamma}(Q^2)\simeq 0.019$ in the $Q^2\sim 9-12$~GeV$^2$ region 
as compared to  $Q^2 F_{\pi^0\gamma^*\gamma}(Q^2)\simeq 0.015$ at $Q^2\sim 4-6$~GeV$^2$, 
reported by BaBar~\cite{BABAR}, are difficult to describe in the framework of 
QCD collinear factorization without invoking unconventional models of the pion DA with 
large end-point enhancements, see Ref.~\cite{I} for several examples and 
the discussion. The new data from BELLE $Q^2 F_{\pi^0\gamma^*\gamma}(Q^2)\simeq 0.016$ at $Q^2\sim 9-12$~GeV$^2$     
are much easier to accommodate within a ``standard'' scenario: e.g the ``holographic'' model inspired by
gauge-gravity correspondence provides a good fit, see above. This discrepancy must be clarified.
Provided the Gegenbauer coefficient $a_2$ is determined from lattice calculations and/or fits to the data
at lower $Q^2$, experimental data at $Q^2\sim 10$~GeV$^2$ will allow one to obtain quantitative information
(constraints) on the shape of the pion DA beyond the value of its second moment. E.g. 
the coefficient $a_4$ in the Gegenbauer expansion can be determined in this way.

\subsubsection*{3.~~large $Q^2 > 20$~GeV$^2$}

The transition form factor is expected to approach the scaling behavior in this region.
To make this statement more quantitative, consider the following ratio:
\begin{equation}
 R_{(40/20)} = 2 \frac{F_{\pi^0\gamma^*\gamma}(Q^2=40\,\mathrm{GeV}^2)}{F_{\pi^0\gamma^*\gamma}(Q^2=20\,\mathrm{GeV}^2)}\,.
\end{equation}
We obtain (cf. Fig.~\ref{fig1:SUMMARY})
\begin{equation}
   1.01 <  R_{(40/20)} < 1.07 \,,
\label{scaling}
\end{equation}
where the lower number corresponds to the BMS model~\cite{BMS} and the upper one to the ``flat'' DA. 
There is, on the other hand, no reason to
expect that the form factor at $Q^2 > 20$~GeV$^2$ is close to the Brodsky-Lepage limit
\begin{equation}
  Q^2 F_{\pi^0\gamma^*\gamma}(Q^2) \stackrel{Q^2\to\infty}{=} \sqrt{2}f_\pi = 0.185
\end{equation}
because the asymptotic behavior is achieved in QCD very slowly. Thus, this value is much
less constrained. For the range of the models shown in 
Fig.~\ref{fig1:SUMMARY} we obtain e.g. 
\begin{equation}
    0.163 \le [Q^2 F_{\pi^0\gamma^*\gamma}(Q^2)]\Big|_{Q^2=30\,\mathrm{GeV}^2} \le 0.245
\label{bound}
\end{equation}
(all numbers in GeV).
Much of the excitement produced by the BaBar experimental data was due to an over-interpretation, 
from our point of view, of the power-law fit to the form factor 
$Q^2 F_{\pi^0\gamma^*\gamma}(Q^2) \simeq 0.182 (Q^2/10)^{0.25}$ provided in Ref.~\cite{BABAR}.
This fit suggests a large scaling violation $R^{\rm BaBar}_{(40/20)} \sim 1.2$ which is significantly 
outside of the range in Eq.~(\ref{scaling}) and difficult to reconcile with 
asymptotic freedom. The fit is, however, dominated by the data at lower momentum transfers 
so that such a rise at large $Q^2$ is not warranted.
The same data can be described as (cf.~\cite{Uehara:2012ag}) 
\begin{eqnarray}
  Q^2 F^{\rm BaBar}_{\pi^0\gamma^*\gamma}(Q^2) &=& A Q^2/(B+Q^2) 
\end{eqnarray}
with the values of the parameters $A= 0.230\pm 0.008$~GeV and $B=2.57\pm 0.31$~GeV$^2$.
Whereas the quality of this fit, $\chi^2=1.73/{\rm d.o.f.}$ is, admittedly, worse than
of the BaBar power-law parametrization, $\chi^2=1.04/{\rm d.o.f.}$, it shows that the evidence
against the scaling behavior is at the level of less than two standard deviations.

Since there are strong theoretical arguments that the scaling violation should not exceed 
a few percent, cf. Eq.~(\ref{scaling}), and since experimental errors are currently much larger, 
we prefer to average the existing data over the $Q^2>20$~GeV$^2$ region using kernel density estimation. 
Taking into account three data points with the largest values of $Q^2$ one obtains
\begin{eqnarray}
  {}[Q^2 F_{\pi^0\gamma^*\gamma}(Q^2)]\Big|^{\rm BaBar}_{Q^2 > 20 \,\mathrm{GeV}^2}
  &=& 0.238 \pm 0.029\,,
\nonumber\\
  {}[Q^2 F_{\pi^0\gamma^*\gamma}(Q^2)]\Big|^{\rm BELLE}_{Q^2 > 20 \,\mathrm{GeV}^2}
  &=& 0.204 \pm 0.026  
\end{eqnarray}  
for BaBar and BELLE experiments, respectively.
We observe that the two numbers are consistent to each other within $1.5\sigma$ and  
are both somewhat larger than the QCD calculation with the asymptotic DA (the lower number in Eq.~(\ref{bound})).
This result supports the common wisdom that the pion DA at low scales is broader than the asymptotic distribution.
A much higher accuracy is needed, however, to discriminate between different models and 
provide quantitative constraints.
Future experimental data in the large-$Q^2$ region would also be very interesting for the comparison 
of time-like and space-like transition form factors which are predicted to be equal at asymptotically 
large momentum transfers. The corresponding discussion goes beyond the tasks of this short note. 
   
{}Finally, we want to make a few general remarks. 
First, one has to be very careful in comparing the statements 
on the shape of the pion DA that are obtained in different theoretical approaches. 
For example, we find, in apparent contradiction to Ref.~\cite{Radyushkin:2009zg}, 
that a ``flat'' pion DA $\phi^{\rm flat}=1$ fails to describe the BaBar data, as it strongly overestimates the form 
factor at moderate $Q^2$, see Fig.~\ref{fig1:SUMMARY}. 
The reason for this disagreement is that in Ref.~\cite{Radyushkin:2009zg} a very large 
(nonstandard) nonperturbative correction is assumed that is beyond the operator product expansion. 
It is this assumption and not the shape of the pion DA that is crucial to obtain a strong scaling violation.

Second, we shortly comment on the relation of our results with those of the Bochum-Dubna group
(for the latest update see~\cite{Bakulev:2011rp}) which are obtained in the same approach. 
One difference is that we do not include NNLO perturbative corrections induced by the running 
coupling and prefer to work consistently to the NLO accuracy. 
The reason is that terms corresponding the the running coupling (the so-called large-$\beta_0$ approximation)
usually strongly overestimate the full perturbative correction. In addition there is danger of double counting 
with twist-four contributions because of QCD renormalons, see \cite{Beneke:1998ui} for a review.  
Another difference is that we use somewhat larger values of the so-called Borel parameter, 
$M^2\simeq 1-2$~GeV$^2$ vs. $M^2\simeq 0.7-0.9$~GeV$^2$ see Ref.~\cite{I}.
Nevertheless,  the numerical difference between our calculations (for a given DA) does not exceed 5-8\% at
$Q^2=4-8$~GeV$^2$, and appears to be negligible at larger momentum transfers. 

Note that in difference to Ref.~\cite{Bakulev:2011rp} we do not have any preferred model of the pion DA.
Our analysis is entirely devoted to the possibility of extraction of the parameters of the DA (Gegenbauer coefficients)
from the experimental data using collinear factorization, in full similarity with the determination of
conventional parton distributions from the data on inclusive reactions.

We do not understand the claim made in Ref.~\cite{Bakulev:2011rp} that parametrizations of the pion DA 
introduced in our work \cite{I} fail to describe the BaBar data: E.g. model~II shown in Fig.~\ref{fig1:SUMMARY} 
leads to $\chi^2 = 1.06$ per data point, where the experimental errors and theoretical uncertainties of the 
calculation for a given DA are added in quadrature. For the central values of the parameters 
and not taking into account theoretical uncertainties we obtain for this model $\chi^2 = 1.34$
which is also acceptable.

To summarize, the measurements reported by the BELLE collaboration~\cite{Uehara:2012ag} are by no means a 
game changer, but should somewhat take the heat off 
theorists struggling to invent a nonperturbative mechanism 
capable to postpone (or even invalidate, in the most drastic scenarios) the onset of QCD factorization in the 
pion transition form factor. The new generation of experimental data that is expected to come from super-B 
factories will settle this question and allow the $\pi^0\gamma^*\gamma$ transition 
to serve its purpose as the gold-plated reaction in the theory of hard exclusive reactions.
We have to warn, however, that this process alone is not sufficient for an unambiguous determination 
of the pion DA: one needs to do a global fit of all available hard exclusive processes 
including pions and eventually also include the lattice data.

\section*{Acknowledgements}
S.A. is grateful to the QCD theory group in Regensburg University for hospitality and
financial support.


\begin{thebibliography}{99}

\bibitem{I} 
  S.~S.~Agaev, V.~M.~Braun, N.~Offen and F.~A.~Porkert,
  Phys.\ Rev.\ D {\bf 83}, 054020 (2011)

\bibitem{Uehara:2012ag} 
  S.~Uehara {\it et al.}  [Belle Collaboration],
  arXiv:1205.3249 [hep-ex].

\bibitem{BABAR}
  B.~Aubert {\it et al.}  [The BABAR Collaboration],
  Phys.\ Rev.\  D {\bf 80} (2009) 052002.


\bibitem{Gronberg:1997fj}
  J.~Gronberg {\it et al.}  [CLEO Collaboration],
  Phys.\ Rev.\  D {\bf 57} (1998) 33.

\bibitem{BMS} For this calculation we use the Gegenbauer coefficients $a_2=0.20$ and 
$a_4=-0.14$ at $\mu^2=1$~GeV$^2$ as quoted in:
  A.~P.~Bakulev, S.~V.~Mikhailov, A.~V.~Pimikov and N.~G.~Stefanis,
  arXiv:1205.3770 [hep-ph].

\bibitem{Brodsky:2007hb} 
  S.~J.~Brodsky and G.~F.~de Teramond,
  Phys.\ Rev.\ D {\bf 77}, 056007 (2008).

\bibitem{Radyushkin:2009zg} 
  A.~V.~Radyushkin,
  Phys.\ Rev.\ D {\bf 80}, 094009 (2009).

\bibitem{Unverzagt:2012zz} 
  M.~Unverzagt,
  J.\ Phys.\ Conf.\ Ser.\  {\bf 349}, 012015 (2012).


\bibitem{Bakulev:2011rp} 
  A.~P.~Bakulev, S.~V.~Mikhailov, A.~V.~Pimikov and N.~G.~Stefanis,
  Phys.\ Rev.\ D {\bf 84}, 034014 (2011)


\bibitem{Beneke:1998ui} 
  M.~Beneke,
  Phys.\ Rept.\  {\bf 317}, 1 (1999).


\end{thebibliography}
\end{document}